\documentclass[a4paper,11pt]{article}
\usepackage{adjustbox}
\usepackage{rotating} 
\usepackage{nameref} 
\usepackage{anyfontsize} 
\usepackage{float} 
\usepackage{graphicx} 
\usepackage{indentfirst} 
\usepackage{tabu} 
\usepackage{epstopdf} 
\usepackage{url} 
\usepackage{hyperref} 
\usepackage[table,xcdraw]{xcolor}
\usepackage{tikz} 
\usepackage[toc,page]{appendix}
\usepackage{threeparttable} 
\usepackage{adjustbox}
\usepackage[table,xcdraw]{xcolor}

\usepackage{hyperref} 
\title{ Not All Browsers Are Created Equal: Comparing Web Browser Fingerprintability }
\author{Nasser Mohammed Al-Fannah \\ Information Security Group \\ Royal Holloway, University of London \\ \href{mailto:nasser@alfannah.com}{nasser@alfannah.com} \and Wanpeng Li \\ School of Mathematics, Computer Science and Engineering\\ City, University of London \\ \href{mailto: wanpeng.li@city.ac.uk}{wanpeng.li@city.ac.uk}}
\begin{document}
	\maketitle
	\begin{abstract}
		Browsers and their users can be tracked even in the absence of a persistent IP address or cookie.  Unique and hence identifying pieces of information, making up what is known as a fingerprint, can be collected from browsers by a visited website, e.g.\ using JavaScript.  However, browsers vary in precisely what information they make available, and hence their fingerprintability may also vary.  In this paper, we report on the results of experiments examining the fingerprintable attributes made available by a range of modern browsers.   We tested the most widely used browsers for both desktop and mobile platforms. The results reveal significant differences between browsers in terms of their fingerprinting potential, meaning that the choice of browser has significant privacy implications.
	\end{abstract}
	
	\section{Introduction}
	
	\textit{Browser fingerprinting} is a technique that can be used by a web server to uniquely identify a platform; it involves examining information provided by the browser, e.g.\ to website-originated JavaScript.  The notion of browser fingerprinting was first discussed by Eckersley \cite{unique}.  Since Eckersley's seminal work, the range and richness of fingerprinting information retrievable from a browser has substantially increased \cite{beast}.  Of course, web cookies and/or the client IP address can be used for the same purposes, but browser fingerprinting is designed to enable browser identification even if cookies are not available and the IP address is obfuscated, e.g.\ through the use of anonymising proxies.
	
	This paper is intended to help understand whether, and to what degree, widely-used browsers vary in the quantity and quality of the fingerprinting attributes they make available.  This would enable us to learn their relative fingerprintability.  We describe a series of systematic tests performed on currently available browsers, which show that some browsers reveal substantially more fingerprinting information than others; hence users of the least privacy-respecting browsers can be more readily be identified and/or tracked.  We performed the tests using a specially established website \url{https://fingerprintable.org}\footnote{All the scripts used in our experiments are publicly available --- see Appendix A.}.  This website does not retain any data recovered from visiting browsers, but simply displays the information that it is able to collect from the currently employed browser.  We hope that this site will be a useful tool in promoting general understanding of the privacy threat arising from browser fingerprinting, and more generally from some of the features provided by today's browsers when executing JavaScript.

	Over the last five or six years, a number of authors have performed detailed studies of the effectiveness of a range of browser fingerprinting techniques (e.g.\ \cite{detective,unique,metrics,Canvas,monster,battery}).  In this paper, we use a selection of known fingerprinting approaches to compare the fingerprintability of widely used web browsers on both desktop and mobile platforms.  Since desktop browsers differ significantly from their mobile counterparts in their capabilities and features (e.g.\ plugins cannot be installed on mobile browsers), we made parallel studies for these two platform types.
	
	The remainder of the paper is structured as follows. We start in section 2 with the methodology used in our experiments. In section 3 we discuss the experimental results, followed by an analysis in section 4. In sections 5 and 6 we discuss methods to maintain privacy while browsing and give concluding remarks.

	\section{Methodology}
	
	We performed our experiments on five of the latest and most widely used platform types.  Specifically, we chose to examine browsers running on Windows 10 and Mac OS X 10.12 (Sierra) for desktop platforms, and Android 7.0 (Nugget), iOS 10.2.1 and Windows 10 Mobile for mobile devices.  Further details of the methodology we employed to examine browser fingerprintability, including the set of browsers we examined, are given below.  Precise details of the versions of operating systems and browsers used are given in Appendix B.
	
	\subsection{Browsers}
	As noted above, given the major functional differences between desktop and mobile browsers, we made parallel studies of the two classes.  For both mobile and desktop platforms, we chose to examine the five most widely used browsers according to netmarketshare.com\footnote{\url{https://www.netmarketshare.com/browser-market-share.aspx} [accessed on 03/03/2017]}. 
	\begin{itemize}
		\item The desktop browsers we examined were \textbf{Chrome, Internet Explorer, Firefox, Edge} and \textbf{Safari}.
		\item The mobile browsers used in our tests were \textbf{Chrome, Safari, Opera Mini, Firefox} and \textbf{Edge}.  We excluded Mobile Internet Explorer and Android Browser because they are no longer being developed or included with new devices. Specifically, Google has replaced its native Android browser with Chrome, and Microsoft has replaced Internet Explorer with Edge in Windows 10 Mobile.
	\end{itemize}
	
	\subsection{ Installation Options}
	The use of add-ons and plugins can both increase and decrease the information available for fingerprinting.  The presence of add-ons inherently increases fingerprinting capabilities, since the set of add-ons (information that is typically available to executing JavaScript) helps individualise a browser; in addition, some add-ons reveal information that can identify the user or browser \cite{counter}.  On the other hand, specially designed anonymizing add-ons can be used to conceal a browser's fingerprint \cite{counter}.  To avoid biasing the results, in our tests we used clean installations of browsers so that they did not include any add-ons or plugins other than those installed and enabled by default.  We could have chosen to disable even those add-ons that are present and enabled by default, but we chose to leave them on the basis that many users will not change the browser default settings; hence testing the browser ``out of the box" gives the fairest assessment of its privacy properties.  
	
	In fact, the browsers we examined come with very few installed and enabled add-ons; Edge and Internet Explorer are the only browsers we tested that come with the Flash plugin installed and enabled by default.  Although Chrome comes with the Flash plugin installed, it is disabled. 
	
	The mobile browsers require various permissions to be set as part of their installation.  In addition, browsers may request extra permissions while executing, depending on the features of a visited website (e.g.\ to request permission to take pictures and record video).  For testing purposes, we did not grant any permissions other than those needed for browser installation.

	\subsection{Experimental Scripts}
	To test the fingerprintability of the selected browsers, a web page containing JavaScript was constructed, intended to be served by our experimental website (\url{https://fingerprintable.org}).  Whenever the website is visited by a client browser, e.g.\ one of those being tested, the scripts in the web page interrogate the browser to learn the values of a set of identifying attributes (as discussed in section \ref{att}).  Technical details of the scripts are provided in Appendix A.  The scripts used in the experiments were largely based on those available in the GitHub open repositories.  
	
\begin{figure}[!t]
		\centering
		\includegraphics [width=0.7\textwidth]{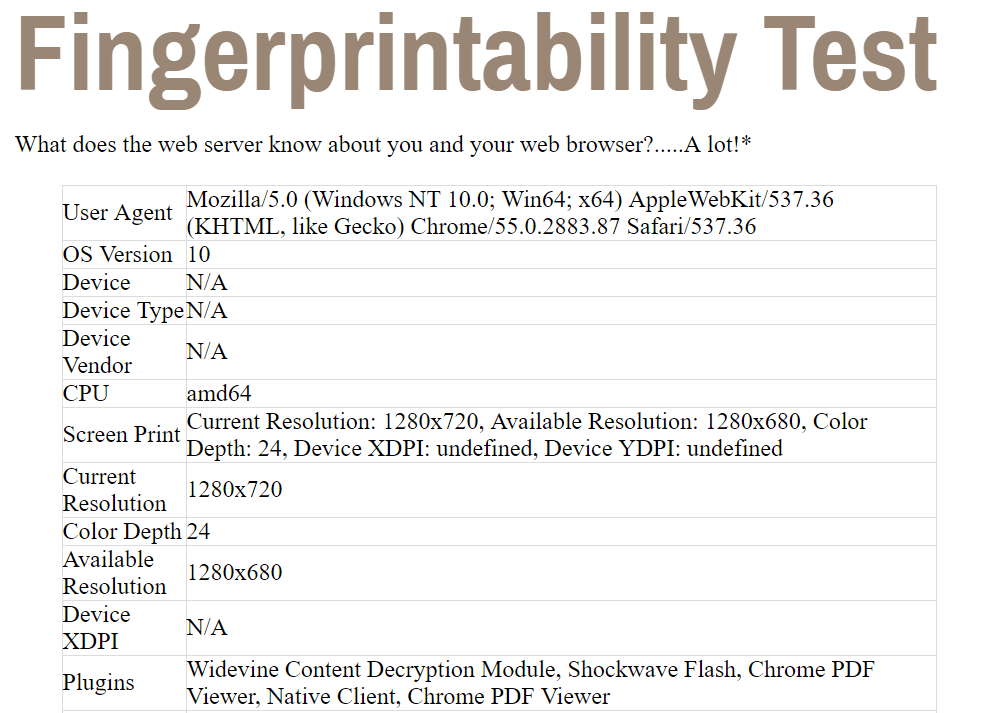}
		\caption{The Fingerprintability page}
		\label{fig:fp}
		
	\end{figure}

	The web page displayed by the browser contains a summary of the fingerprinting information gathered by the script, and thereby provides an instant summary of the privacy properties of the browser.  As mentioned elsewhere, this site is publicly available, and is open for general use.  A partial screenshot of a typically displayed page is shown in Figure \ref{fig:fp}.
	
	The total size of the script used is approximately 70kB; in informal tests it loaded and displayed the results without any noticeable delay.
	\subsection{Attributes} \label{att}
	The original goal of our experiments was to sample all the attributes that can be collected from a web browser.  Any attribute that is not fixed for all browsers has potential value for fingerprinting.  However, a large number of attributes have Boolean values (e.g.\ Java installed?) or one of a very limited set of values (e.g.\ Java version) and hence they typically give relatively little identifying information.  Given the significant number of such attributes, we therefore omitted such attributes from our tests, and focused on those that have the potential to give significantly more information.  
	
	We also omitted attributes that, according to Laperdrix et al.\ \cite{beast}, take more than a few seconds to collect (e.g.\ font metrics \cite{metrics}), or are unreliable for fingerprinting purposes (e.g.\ battery level \cite{battery}).  Additionally, we omitted attributes that are made available by all tested browsers as part of their typical functionality (e.g.\ screen resolution).  It is worth noting that some attributes are related to the user's machine and thus can be used to help identify a specific platform even if a user subsequently switches browsers \cite{cross}.  Others are browser-specific, and hence can only be used for fingerprinting as long as the same browser is used.  
	
	We next discuss in detail the six fingerprinting attributes used in our tests.

	\subsubsection{Fonts through Flash}
	If the Adobe Flash plugin is installed and enabled, it can be used to reveal the set, and installation order, of fonts installed on the user platform; this is known to be a highly discriminating attribute (see, for example, Eckersley \cite{unique}).   Moreover, this attribute can be used to fingerprint a platform even if multiple browsers are used.  However, of the desktop browsers we examined, only Edge and Internet Explorer have Flash installed and enabled by default.  In this respect, Edge is therefore significantly less privacy-protecting than its competitors, since learning the set of installed fonts without using Flash is non-trivial.
	
	None of the mobile browsers we examined support Flash, so the set of installed fonts is not used when comparing the fingerprintability of this class of browsers.  Furthermore, the most widely used mobile OSs (i.e.\ Android and iOS) do not give the user the option to install fonts.
	
	There is another, albeit less accurate, method of discovering the set of installed fonts using website-supplied JavaScript \cite{monster}.  However, we do not consider it as part of our comparison since it works in the same way for all the tested browsers.
	\subsubsection{Device ID(s)}
	A device ID\footnote{ \url{https://browserleaks.com/webrtc\#webrtc-device-id} [accessed on 03/03/2017]} is a hash value generated by a browser by applying a cryptographic hash function to the unique ID of a hardware component in the user platform (combined with other data values); it is retrieved by requesting the WebRTC hardware ID attribute.  WebRTC\footnote{ \url{https://webrtc.org} [accessed on 03/03/2017]} is a set of communications protocols and APIs that provides browsers and mobile applications with Real-Time Communications (RTC) capabilities via simple APIs.  
	
	The main intended application of such device IDs would appear to relate to managing multimedia content, and the platform components whose identifiers are used are typically the loudspeaker, microphone and/or camera.  Since the device ID is computed on other data in addition to a unique hardware identifier, the value computed by a browser will typically change when accessed by different websites.  However, for a single website, the device ID appears likely to remain constant (at least for some browsers) across multiple visits, giving it high value for fingerprinting purposes.   Moreover, a device ID seems to be constant when queried in different ways; for example, we obtained the value via an iframe on a different website, and it gave the same value as that for the framed site.
	
	To the authors' knowledge, the use of this attribute for browser fingerprinting has not previously been discussed, and so its robustness and usefulness for this purpose has yet to be determined.  However, experiments conducted as part of this research show that it has great promise for use in fingerprinting.  Nonetheless, further study is needed to investigate this in greater detail.  Gaining a better understanding of how exactly the device ID is computed by the various browsers would certainly help in such an investigation, although such information does not appear to be publicly available.

	\subsubsection{Canvas Image}
	The Canvas API is a recently introduced HTML5 API that allows websites to render an image for display by the user browser, an alternative to the commonly used technique of downloading an image file from the server \cite{html5}.  Several studies \cite{forget,beast,Canvas} have demonstrated the possibility of uniquely fingerprinting browsers and their host platforms based on subtle differences in how an image is rendered by the browser.   We based our tests on the particularly effective Canvas image fingerprinting approach due to Englehardt and Narayanan \cite{tracking}.  
	
\begin{figure}[!t]
		\centering
		\includegraphics[width=2.2in]{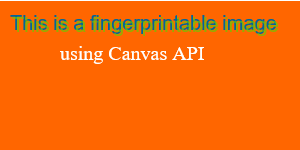}
		\caption{Sample Canvas image used in our tests}
		\label{fig:Canvas}
		
	\end{figure}

	The Canvas API allows the server to request the return of certain details of a rendered image (e.g.\ the RGBA\footnote{red green blue alpha (opacity)} values of rendered image pixels) \cite{Canvas}.  Since each browser appears to have its own rendering algorithm, the returned image will vary depending on the browser as well as the computing environment \cite{Canvas}. A simple means of using this fact for fingerprinting is to hash the returned image details and use this hash value as an attribute.  All the tested browsers rendered the sample Canvas image provided by the test script (see figure \ref{fig:Canvas}), and as a result give a fingerprint for that browser.  Although not part of our comparative experiments, it is interesting to observe that the Tor browser (a modified version of Firefox that uses the Tor network) displays a warning and asks the user for permission before rendering a Canvas image.  However, none of the tested browsers made such a request.
	
	Not only does this attribute enable fingerprinting based on the browser in use, but in some cases it provides the ability to discriminate between two similar platforms running the same browser.  That is, in some cases the image rendered by the same browser will differ given a small change in the computing environment.

	\subsubsection{WebGL Renderer}
	Some browsers provide access to the identity of the vendor and the specific model of the user platform's Graphics Processing Unit (GPU).   These two pieces of information are obtained by requesting the following WebGL attributes: UNMASKED VENDOR WEBGL and UNMASKED RENDERER WEBGL.  These attributes could reveal the Central Processing Unit (CPU) type if there is no GPU or if the GPU is not used by the browser.  
	
	We found that the UNMASKED VENDOR WEBGL either states the browser vendor or the CPU/GPU vendor.  In both cases it does not provide any useful information that cannot be readily found from the UNMASKED RENDERER WEBGL (i.e.\ identifying a vendor is trivial once the full CPU/GPU model details are known) or the \textit{user agent} header (see \ref{ua}) which reveals the browser vendor.  We shall therefore focus solely on the WebGL renderer.

	\subsubsection{User Agent} \label{ua}
	
	This attribute consists of a data string included in the header of HTTP response packets that gives information related to the browser, including its type and version \cite{beast} (e.g.\ \textit{Mozilla/5.0 (Windows NT 10.0; WOW64; rv:50.0) Gecko/20100101 Firefox/50.0}).  It is useful in enabling websites to tailor content to meet the needs of differing browsing platforms.  Whilst a rich source of fingerprinting information, all desktop browsers provide much the same level of detail, and so this attribute is not useful in comparing their fingerprintability; we therefore do not use this attribute when comparing desktop browsers.  However, it remains useful for comparing the fingerprintability of mobile browsers, since in some browsers it includes an indication of the model of the mobile phone.

	\subsubsection{Private IP Address(s)}
	The local IP address of a user platform can be discovered if the executing JavaScript is able to ping a STUN server\footnote{ A STUN server (i.e.\ a Session Traversal of User Datagram Protocol Through Network Address Translators (NATs) server) allows a NAT client to set up interactive communications such as a phone call to a VoIP provider hosted outside the local network.}.   This possibility was apparently first observed by Roesler\footnote{https://diafygi.github.io/webrtc-ips/}.  This fingerprinting technique works for browsers that support WebRTC \cite{tracking}.   Potentially, the revealed IP address(es) could include the client's local IPv4 address as well as one, or more, of the client's unique local addresses (ULAs)\footnote{The ULA is the approximate IPv6 counterpart of the IPv4 private address;  see \url{https://tools.ietf.org/html/rfc4193} [accessed 03/03/2017]} thus making it more fingerprintable. 
	
	Ideally, a website cannot discover the real public IP address of a user platform that is employing a VPN\@.  However, as discussed by Perta et al\@. \cite{vpn}, a website can learn the public IP address as well as the IP address assigned by NAT or VPN of a visiting browser by exploiting a feature of its WebRTC  implementation \cite{augment}.  However, it was previously reported \cite{vpn} that an IP address leak does not occur in all VPN implementations.

	\subsection{Performing the Experiments}
	Platforms of the specified types, running the chosen operating systems, were equipped with the relevant browsers (clean installs, as discussed above).  The browser was then made to visit the test website (\url{https://fingerprintable.org}) and the data generated by the script was collected and recorded.  The 10 datasets (five for the desktop platform and five for the mobile platform) generated were then processed and used to derive the information given in section \ref{ch5:results} below.
	
	\subsubsection{Attribute Processing}
	Each browser was tested for the retrievability of discriminating information for each of the six fingerprinting attributes described in section \ref{att}.   For most attributes, it was straightforward to determine whether or not the browser returned any fingerprintable values.  However, some attributes required some processing to be useful.  For example, an attribute such as User Agent always returns a string of information.  The key difference between one browser and another was whether it included information specific to the system hosting it.  These differences were observed and noted.  
	
	The device ID was tested for both its existence as well as its persistence.  We observed that the browsers that calculate such a value, at some point calculate a new value.  The main difference between browsers in this respect is in the nature of the trigger that causes recalculation.  This means that some browsers have a more persistent device ID, i.e.\ one that is more valuable for fingerprinting, than others.
	
	The Canvas image typically returns the same hash value when tested on identical platforms.  Some browsers also give the same hash value when running on two devices that have relatively similar specifications.   Such rendering makes the attribute less useful for fingerprinting.  To find these cases we tested and compared Canvas rendered images of each browser on two devices with similar hardware (for device specifications see Appendix C).
	
	\subsubsection{Fingerprintability Index}
	For comparison purposes, we rank each attribute as having a \textit{high} (3), \textit{medium} (2) or \textit{low} (1) \textit{Fingerprintability Index} (FI), where higher FI indicates an attribute giving more information useful for fingerprinting.  These assignments are based on previous work as well as our own qualitative estimations.  We have refrained from using the term \textit{entropy} or precise entropy values taken from the prior art, as values are not available for all the attributes we consider in our study.  It is important to note that, regardless of the ranking of attributes, all attributes in our study provide relatively high entropies, as explained earlier in the paper.
	
	The fonts attribute is ranked as \textit{high} as it is a highly discriminating piece of information \cite{unique}.   Device IDs also have the potential of being highly discriminating; however, as discussed earlier, browsers that provide device IDs differ in terms of the persistence of the values.  This attribute is therefore assigned \textit{high} if the browser shows no signs of changing this value under typical browser usage, and is assigned \textit{medium} if a browser provides a new value with every browsing session.  It is assigned \textit{low} if a browser provides a new value with every visit or page refresh.  
	
	We rank the Canvas API attribute as \textit{medium}, based on the analysis of Laperdrix et al.\ \cite{beast}.  However, we as rank it as \textit{low} for any browser that returns the same image hash value on two devices with similar specifications.  The WebGL information is ranked as \textit{low}, as Alaca et al.\ \cite{augment} argue that it provides relatively little information useful for fingerprinting.  This is expected since many devices could be using an identical CPU and/or GPU.
	
	The user agent string reveals a lot of information valuable for fingerprinting \cite{augment, cross, beast}.  However, we rank it only as \textit{medium} since we focus here purely on whether or not it includes information on the mobile phone model.  We assign a rank of \textit{low} to the leaking of private IP addresses.  This is because most clients are assigned local IPv4 addresses in the 192.168.0.x range \cite{augment} and such IP addresses tend to be dynamically assigned and so can change regularly.  However, we assign a \textit{medium} ranking to any browser that reveals one, or more, of the client's ULAs in addition to the aforementioned IPv4 address.
	
\section{Results} \label{ch5:results}
We next summarize the results of our experiments.  We divide the discussion into two parts, first addressing the tests on desktop platforms and second the experiments using mobile devices.

\subsection{Desktop Browsers}
\subsubsection{Overview}
We summarize below the key observations arising from our examination of desktop browsers.
\begin{itemize}
	\item{\textbf{Chrome}}
	did not reveal the set of installed fonts through Flash probing despite the presence of the Flash plugin; this was because the plugin is disabled by default.  Chrome is unique in generating a very discriminating device ID\@.  The value remained the same for at least a month, and seems unlikely to change until the browser cache is cleared.  
	
	Canvas image rendering in Chrome resulted in the same hash value on both test machines.  We made the same observation on the mobile version of Chrome.
	
	When probed for the WebGL attribute values, Chrome gave the \textit{full} details of the GPU, including the name and version of the installed graphics API.  The local IPv4 and temporary IPv6 addresses were also revealed.
	
	\item 
	Because \textbf{Internet Explorer} comes with the Flash plugin installed and enabled by default, it can be used to determine the set of installed fonts.  Internet Explorer does not disclose any device IDs (due to its lack of WebRTC support).  The hash value of the image produced using the Canvas API was the same as that generated by Edge on both test machines.  Internet Explorer revealed the specific model of the CPU\@.  However, Internet Explorer did not reveal the local IP address.   
	\item {\textbf{Firefox}} does not include the Flash plugin, and hence it does not reveal the list of installed fonts through Flash probing.  It generated device IDs, although this attribute is less discriminating than in Chrome as the device IDs change with every browser session.  
	
	Firefox produced two different hash values for the canvas-rendered images on the two test machines.  The WebGL probing yielded \textit{Mozilla} as both the vendor and renderer (i.e.\ neither CPU or GPU were revealed).  However, Firefox revealed the client's local IPv4 address.
	
	\item {\textbf{Safari}} revealed no information for most of the attributes we tested.  This is mainly because of its lack of full WebRTC support and the absence of the Flash plugin.  However, it does support the Canvas API and produced the same image hashes on the two test devices.  Safari also revealed the WebGL renderer details.
	
	\item
	Just like Internet Explorer, \textbf{Edge} comes with the Flash plugin installed and enabled by default.  This reveals the set of fonts installed on the computer, which is highly valuable for fingerprinting.  Edge generates device IDs but they change every time a website is revisited or even refreshed.  It gave the same canvas hash value as Internet Explorer on the test machines.  It also revealed the CPU model.  Moreover, amongst tested desktop browsers, it was unique in exposing three IP addresses, namely the client's local IPv4, IPv6 and ULA.
\end{itemize}

\begin{table}[!htb]
	{\fontsize{9}{10}\selectfont
	\centering
	\caption{Desktop browser fingerprintability}
	\label{ch5:desktop}
	
	\begin{tabular}{|c|c|c|c|c|c|}
		\hline
		\rowcolor[HTML]{ EFEFEF} 
		Attribute / Browser & Chrome & Internet Explorer & Firefox & Safari & Edge \\ \hline
		Fonts 	& 
		- 	& 
		\textbullet \textbullet \textbullet	& 
		- 	& 
		-	& 
		\textbullet \textbullet \textbullet \\ \hline
		Device ID& 
		\textbullet \textbullet \textbullet & 
		- & 
		\textbullet \textbullet & 
		- &
		\textbullet
		\\ \hline
		Canvas & 
		\textbullet & 
		\textbullet &
		\textbullet \textbullet &
		\textbullet &
		\textbullet 
		
		\\ \hline
		
		WebGL Renderer & 
		\textbullet \textbullet	&
		\textbullet	&
		-	&
		\textbullet	&
		\textbullet \\ \hline
		
		Local IP Address& 
		\textbullet \textbullet &
		-	&
		\textbullet &
		- &
		\textbullet \textbullet	\textbullet
		\\ \hline

		\rowcolor[HTML]{ EFEFEF} 
		Fingerprintability Index & 
		\textbf{8} & 
		\textbf{5} & 
		\textbf{5} & 
		\textbf{2} & 
		\textbf{9} \\ \hline
		
	\end{tabular}
	\begin{tablenotes}\footnotesize
		
		\item \textbullet = low; \textbullet \textbullet = medium; \textbullet \textbullet \textbullet = high
	\end{tablenotes}	
	}
\end{table}

\subsubsection{Discussion}
The results of our tests are summarized in Table \ref{ch5:desktop}.  Only Chrome, Firefox, and Edge provided device IDs.  The fingerprintability of this attribute varies significantly between tested browsers.  Chrome device IDs are consistent and do not change unless the user selects the \textit{private browsing mode}\footnote{Chrome did not assign a device ID when private mode was enabled.} feature or clears the browser cache.  The Firefox device ID remained the same during multiple visits in a single browsing session, but changed once the browser was reopened.  Of the browsers generating device IDs, Edge gave the value that changed most readily; merely refreshing a web page caused Edge to generate a new value.  This makes this attribute in Edge of very limited use for fingerprinting.

All the tested browsers support the Canvas API and rendered the scripted image in our test, i.e.\ they all reveal this fingerprinting attribute.  However, in the case of Firefox, the image resulted in a different hash when rendered on the two test machines.  As a result, the canvas-rendered images attribute is more fingerprintable in Firefox than the other tested browsers.

With the exception of Safari and Internet Explorer, all the tested browsers exposed the client's local IPv4 address.  Both Edge and Chrome also revealed the IPv6 address.  However, Edge was the only tested browser to reveal the client's ULAs.  Overall, Edge was the most fingerprintable (FI: 9) and Safari the least (FI: 2).

\subsection{Mobile Browsers}
\subsubsection{Overview}
Summarized below are the main observations arising from our examination of mobile browsers.
\begin{itemize}
	
	\item
	The \textbf{Chrome} user agent revealed the specific phone model.  Just like its desktop counterpart, Chrome provided persistent device IDs. Chrome's rendering of the canvas image resulted in the same hash on both testing devices.  It also revealed the vendor and model of the GPU, as well as the local IPv4 and ULA addresses.
	\item
	\textbf{Safari} mobile did not reveal much fingerprinting information except the information derivable from rendering the canvas image and the CPU model through the WebGL API.  It rendered the same canvas image on both test devices.
	
	\item The \textbf{Opera Mini} user agent revealed the phone model. It provided device IDs that were similar to Firefox in terms of calculating a new value with every new browsing session.  It also rendered unique canvas images on tested devices. Moreover, it revealed the GPU model, as well as the local IPv4 and ULA addresses.
	
	\item
	\textbf{Firefox} did not reveal the phone model in the user agent field, which makes this attribute significantly less revealing.  However, Firefox did provide device IDs in the same way as its desktop counterpart.  It also rendered unique canvas images on tested devices and allowed the retrieval of the client's local IPv4 and ULA addresses.  The WebGL did not reveal the vendor nor renderer.
	
	\item
	The \textbf{Edge} user agent included the model of the phone.  Edge provided device IDs but, like its desktop counterpart, the IDs change with every page refresh or revisit.  It also revealed the model of the GPU\@.  The canvas-rendered image was the same on both test devices.  Unlike its desktop version, Edge did not expose any private IP addresses.
\end{itemize}

\begin{table}[!htb]
	{\fontsize{9}{10}\selectfont
	\centering
	\caption{Mobile browser fingerprintability}
	\label{ch5:mobile}
	
	\begin{tabular}{|c|c|c|c|c|c|}
		\hline
		\rowcolor[HTML]{ EFEFEF} 
		Attribute / Browser & Chrome & Safari & Opera Mini & Firefox & Edge\\ \hline
		User Agent & 
		\textbullet \textbullet &
		-	&
		\textbullet \textbullet &
		-	&
		\textbullet \textbullet
		\\ \hline
		Device ID& 
		\textbullet \textbullet \textbullet &
		-	&
		\textbullet \textbullet &
		\textbullet \textbullet &
		\textbullet 
		
		\\ \hline
		Canvas & 
		\textbullet &
		\textbullet &
		\textbullet \textbullet &
		\textbullet \textbullet &
		\textbullet
		\\ \hline
		
		WebGL Renderer & 
		\textbullet &
		\textbullet &
		\textbullet &
		-&
		\textbullet

		\\ \hline
		Local IP Address& 
		\textbullet \textbullet &
		-	&
		\textbullet \textbullet &
		\textbullet \textbullet &
		-	
		\\ \hline

		
		\rowcolor[HTML]{ EFEFEF} 
		Fingerprintability Index & 
		\textbf{9} & 
		\textbf{2} & 
		\textbf{9} & 
		\textbf{6} & 
		\textbf{5} \\ \hline
		
	\end{tabular}
	
	\begin{tablenotes}\footnotesize
		
		\item \textbullet = low; \textbullet \textbullet = medium; \textbullet \textbullet \textbullet = high
	\end{tablenotes}	
		}
\end{table}

\subsubsection{Discussion}

The results of our tests are summarized in Table \ref{ch5:mobile}.  Chrome, Opera Mini and Edge included the phone model as part of the user agent field.  With exception of Safari, all tested browsers calculated device IDs.

Although all tested browsers rendered the canvas image, Chrome, Safari, and Edge (both desktop and mobile) rendered exactly the same image on the test devices with similar specifications.  This makes Chrome, Safari and Edge canvas-rendered images less fingerprintable than the other tested browsers.

Chrome, Opera Mini, and Firefox exposed the local IPv4 addresses.  However, unlike their desktop counterparts, they also exposed the client's ULA(s).  Overall, Chrome and Opera Mini were the most fingerprintable browsers (FI: 9).  Just like its desktop counterpart, Safari was the least fingerprintable (FI: 2).
\subsection{Other Remarks}
It seems reasonable to expect that browser fingerprinting based on the Flash plugin will soon become irrelevant given the imminent disappearance of Flash \cite{beast}.  In regards to the Canvas API, it is important to note that it is not the rendering aspect of the Canvas API that endangers user privacy but the ability to retrieve details of the rendered image by visited websites.  Thus, if this feature was removed from the Canvas API, it would eliminate any possible fingerprinting based on it (at least using current methods).

Device IDs have the potential to seriously endanger user privacy, especially given their persistence in Chrome.  Moreover, Chrome's persistent device IDs seem unnecessary, given that Edge constantly provides new values.

	\section{Circumventing Fingerprinting}
	
	A number of authors have considered the problem of reducing the degree to which a browser can be fingerprinted (e.g.\ \cite{unique,counter}).  Indeed, a range of tools exist which are designed to make browsing more anonymous \cite{detective,lies}.  We, therefore, do not explore this topic in detail, but simply mention four simple measures that can be employed to reduce the usefulness of the fingerprinting attributes we studied in this paper.
	\begin{itemize}
		\item{\textbf{Disable the Canvas API}}
		Despite significant variations in the amount of information that can be collected depending on the browser, the Canvas fingerprint is supported by them all.  This fingerprint alone can make any browser highly fingerprintable.  Currently, Canvas support in the tested browsers can be blocked by specialised add-ons such as \textit{CanvasBlocker}\footnote{\url{https://addons.mozilla.org/en-gb/firefox/addon/canvasblocker}}. 
		\item{\textbf{Disable Flash}}
		The number of websites that use Adobe Flash is reducing, and most web browsers are discontinuing support for it \cite{beast}.  So, anyone using a desktop browser that has the Flash plugin installed and enabled may wish to consider disabling it.  This will prevent a website using Flash to discover the installed fonts, and the order in which they were installed.
		\item{\textbf{Disable WebRTC}}
		This feature is relatively new, and the discovery of security vulnerabilities, such as those discussed in this paper, is perhaps to be expected.  Disabling WebRTC would prevent a website from easily retrieving a client's local IP address(es) or public IP address when using a VPN.  Disabling WebRTC is typically possible through the browser user settings.
		\item{\textbf{Anonymizing Add-ons}}
		There are many anonymizing add-ons available for browsers (e.g.\ \textit{Privacy Badger}\footnote{{\fontsize{7}{7}\selectfont\url{https://chrome.google.com/webstore/detail/privacy-badger/pkehgijcmpdhfbdbbnkijodmdjhbjlgp}}} and \textit{NoScript}\footnote{\url{https://addons.mozilla.org/en-gb/firefox/addon/noscript/}}) that reduce or disguise a browser fingerprint.  These add-ons, and others, have been tested and discussed in detail by Fiore et al.\ \cite{counter}.
	\end{itemize}
	\section{Conclusions}
	Our tests have investigated an aspect of browser fingerprinting that has not previously been explored in literature, namely looking at the differences between browsers in terms of the amount of information they reveal.  Some mobile browsers seem to unnecessarily give out the specific phone model.  Moreover, WebRTC has introduced several privacy-compromising properties that need to be revisited.  Increasing numbers of browsers support WebRTC, and so, unless the issues with it are addressed, browser fingerprintability seems set to increase.  Rendering images via the Canvas API provides a very discriminating fingerprinting attribute, and all tested browsers support it.  It would therefore be highly desirable if all browsers asked for user permission before rendering a Canvas image, or at least disabled the option that allows servers to retrieve details of the rendered image.
	
	Users concerned about their traceability via fingerprinting should also consider selecting their browser with our results in mind.  At the time we performed our experiments, Safari would appear to be the best choice in this respect on both mobile and desktop platforms.  Despite Chrome being the most widely used browser, it proved to be one of the most fingerprintable.

	\bibliographystyle{plain}
	\bibliography{fp}
	
	\begin{appendices}
		\section{Test Code} 
		The scripts used in our experiments were gathered from the following websites:
		\begin{itemize}
			\setlength{\itemsep}{0pt}
			\setlength{\parskip}{0pt}
			\setlength{\parsep}{0pt}  
			\item https://clientjs.org \\
			\item https://github.com/spleennooname/GLeye \\
			\item https://github.com/muaz-khan/DetectRTC

		\end{itemize}
		Some scripts were modified to suit our testing.  All the code we used for testing is available at our website \url{https://fingerprintable.org}.
		{\fontsize{8}{10}\selectfont
			
			\section{Browser and OS Versions} 
			{
				\centering
				\label{browsers}
				\begin{tabular}{|c|c|}
					\hline
					\rowcolor[HTML]{ EFEFEF} 
					Browser & OS 
					\\ \hline
					\multicolumn{2}{|c|}{\textbf{Desktop}}
					\\ \hline
					Chrome 56.0.2924.87 (64-bit) & Windows 10.0.14393 Build 14393
					\\ \hline
					Microsoft Internet Explorer 11.576.14393.0 & Windows 10.0.14393 Build 14393
					\\ \hline
					Firefox 51.2 (32-bit) & Windows 10.0.14393 Build 14393
					\\ \hline
					Microsoft Edge 38.14393.0.0 & Windows 10.0.14393 Build 14393
					\\ \hline
					Safari 10.0.3 (12602.4.8) & macOS Sierra 10.12.3
					\\ \hline
					\multicolumn{2}{|c|}{\textbf{Mobile}}
					\\ \hline
					Chrome 56.0.2924.87 & Android 7.0 (Build 39.2.A.0.374)
					\\ \hline
					Safari 602.1 & iOS 10.2.1(14d27)
					\\ \hline
					Opera Mini 22.0.2254.113472 & Android 7.0 (Build 39.2.A.0.374)
					\\ \hline
					Firefox 51.0.3 & Android 7.0 (Build 39.2.A.0.374)
					\\ \hline
					Microsoft Edge 38.14393.693.0 & Windows 10 Mobile (OS Build: 10.0.14393.693)
					\\ \hline
				\end{tabular}
			}

			\section{ Specifications of Devices Used for Experiments }
			\centering
			\begin{adjustbox}{center}
				\begin{tabular}{|c|m{16em}|m{13em}|m{4em}|}
					\hline
					\rowcolor[HTML]{ EFEFEF} 
					OS & CPU & GPU & RAM
					\\ \hline
					\multicolumn{4}{|c|}{\textbf{Desktop}}
					\\ \hline
					Windows & Intel Core i7-4720HQ 2.6GHz & NVIDIA GeForce GTX 960M & 16.0 GB 
					\\ \hline
					Windows & Intel Core i5-5200U 2.2GHz & Intel HD Graphics 5500 & 12.0 GB
					\\ \hline
					macOS & Intel Core i5 2.7GHz & Intel Iris Graphics 6100 & 8.0 GB
					\\ \hline
					macOS & Intel Core i7 2.7GHz & Intel HD Graphics 530 & 16.0 GB
					\\ \hline
					\multicolumn{4}{|c|}{\textbf{Mobile}}
					\\ \hline
					Android & Qualcomm Snapdragon 820 64-bit & Adreno 530 & 3.0 GB
					\\ \hline
					
					Android & Qualcomm Snapdragon 801 2.5GHz & Adreno 330 & 3.0 GB
					\\ \hline
					iOS & A8 chip 64-bit & PowerVR GX6450 & 1.0 GB
					\\ \hline
					iOS & A9 chip 64-bit & PowerVR GT7600 & 2.0 GB
					\\ \hline
					Windows & Qualcomm Snapdragon 400 1.2GHz & Adreno 305 & 1.0 GB
					\\ \hline
					Windows & Qualcomm Snapdragon 200 1.2GHz & Adreno 302 & 1.0 GB
					\\ \hline

				\end{tabular}
			\end{adjustbox}
		}

	\end{appendices}
\end{document}